ARTICLE   OPEN 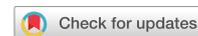

# Experimental implementation of precisely tailored light-matter interaction via inverse engineering

Ying Yan[1,2 ✉], Chunyan Shi[3,5], Adam Kinos[3], Hafsa Syed[3], Sebastian P. Horvath[3,6], Andreas Walther[3], Lars Rippe[3], Xi Chen[4 ✉] and Stefan Kröll[3 ✉]

Accurate and efficient quantum control in the presence of constraints and decoherence is a requirement and a challenge in quantum information processing. Shortcuts to adiabaticity, originally proposed to speed up the slow adiabatic process, have nowadays become versatile toolboxes for preparing states or controlling the quantum dynamics. Unique shortcut designs are required for each quantum system with intrinsic physical constraints, imperfections, and noise. Here, we implement fast and robust control for the state preparation and state engineering in a rare-earth ions system. Specifically, the interacting pulses are inversely engineered and further optimized with respect to inhomogeneities of the ensemble and the unwanted interaction with other qubits. We demonstrate that our protocols surpass the conventional adiabatic schemes, by reducing the decoherence from the excited-state decay and inhomogeneous broadening. The results presented here are applicable to other noisy intermediate-scale quantum systems.

npj Quantum Information (2021) 7:138 ; https://doi.org/10.1038/s41534-021-00473-4

## INTRODUCTION

In the past decades, controlling quantum states is a matter of key importance for state-of-the-art quantum science and technologies, i.e. quantum metrology, quantum computing, and quantum information processing[1–3]. More specifically, precise state control in solid-state systems is primarily limited by decoherence, and thus, performing fast operations is an important mitigation strategy to overcome this limitation. At the same time, operations are also constrained by other characteristics of the system or of the process, such as inhomogeneous variations or fluctuations in driving frequency and power, disturbing interactions with the qubits that are not of interest, etc. In order to compensate for these physical imperfections, engineered operations are required, where speed, robustness, and minimization of disturbances have to be combined. Adiabatic passage[4–7] is one such technique that has been used to make operations robust. It is built on the adiabatic theorem where the state evolves relatively slowly[8,9], such that the robustness is achieved at the cost of speed.

The concept of "shortcuts to adiabaticity" (STA)[10,11] originally proposed to accelerate the slow adiabatic passages, provides fast nonadiabatic routes toward the state preparation or transfer. The STA techniques, counter-diabatic driving (quantum transitionless algorithm)[12–14], and the inverse engineering based on Lewis–Riesenfeld (LR) invariants[15,16], were developed for different proposals. Although these two are proven to be mathematically equivalent, they are quite different in terms of physical implementations[15]. For instance, counter-diabatic driving implies the supplementary interaction to compensate the diabatic transition for speeding up the adiabatic processes in different quantum systems, such as Bose–Einstein condensates in accelerated optical lattices[17], single or ensemble nitrogen-vacancy centers in diamond[18–20], cold atoms[21], and superconducting qubits[22–24]. Meanwhile, inverse engineering based on LR invariants has been utilized for experiments on atomic cooling in a harmonic trap with larger flexibility[25]. In the context of inverse engineering, the pulse engineering is based on an ansatz satisfying the boundary conditions, which gives the user the flexibility to optimize the shortcut protocols in combination with perturbation theory and optimal control for specific experimental requirements[26,27], such as pulse shapes, physical constraints, and various types of noise and errors. This motivates us to explore the optimization of STA by using inverse engineering to achieve the fast and robust quantum control of solid-state qubit in a rare-earth ions (REI) system, by meeting the realistic physical requirements, e.g., to improve the fidelity within the frequency channel and avoid the unwanted transitions outside the qubit frequency channel.

Here, we experimentally implemented the inverse engineering STA based on the LR invariants and used it to create quantum states that have a high fidelity within an engineered frequency interval designated to the qubit operations, while simultaneously having little impact on frequency channels outside of this interval. Instead of the unfeasible or complicated counter-diabatic driving, inverse engineering provides us the required degrees of freedom to engineer the qubit control. We performed an experimental demonstration of the theoretical scheme[28] in a REI solid-state system, which is well-suited for quantum information due to the excellent optical and spin coherence properties[29–31], thus is a competitive approach to quantum computing, quantum memories, and quantum communication. While scalable REI quantum computing requires single instance qubits[32], general properties of REI qubits can still be demonstrated and investigated using ensemble-based qubits. In fact, the ensemble-based REI system may be particularly favorable for demonstrating the capacity to engineer custom solutions to operations, since the qubit is inhomogeneously distributed within a frequency range (here 170 kHz), while other ions that should not be affected exist further out in frequency (a few MHz). Quantum superposition states have previously been prepared with an average fidelity of 93% in this system[5], limited by the dephasing of the optical state, which is transiently used for gate operations. In the current paper, we use the non-adiabaticity to cut down the time spent in the excited state by about a factor of five. This improves the fidelity to between 97 and 98%, although it was still limited by the optical coherence time of the Praseodymium (Pr) species that was used. Switching to another rare-earth species with longer optical coherence time and larger spin level separations, for example, Europium[29,31], should enable fidelities above 99.7% according to simulations.

[1]School of Optoelectronic Science and Engineering & Collaborative Innovation Center of Suzhou Nano Science and Technology, Soochow University, Suzhou, China. [2]Key Lab of Advanced Optical Manufacturing Technologies of Jiangsu Province & Key Lab of Modern Optical Technologies of Education Ministry of China, Soochow University, Suzhou, China. [3]Department of Physics, Lund University, Lund, Sweden. [4]Department of Physical Chemistry, University of the Basque Country UPV/EHU, Apartado 644, Bilbao, Spain. [5]Present address: Zurich Instruments, Technoparkstrasse 1, Zurich, Switzerland. [6]Present address: Department of Electrical Engineering, Princeton University, Princeton, NJ, USA. ✉email: yingyan@suda.edu.cn; chenxi1979cn@gmail.com; stefan.kroll@fysik.lth.se

Creating superposition states with higher fidelities than previously reached in REI systems also requires a better control of the quantum state tomography (QST) pulses used to analyze the created superposition states. This leads us to unravel imperfections in the tomography operations that yielded different readout fidelities for different target states. This effect was previously not known, and we have found the cause for these state differences to be related to mixing with nonresonant levels of the ions. By accurate simulation of the full six-level quantum system, we characterize these issues, which we expect will be useful for all future gate experiments in any system where there are potential complications due to off-resonant excitation of nearby quantum states.

Overall, our experiments prove that the combination of LR invariants shortcut passage and optimization of pulse parameters is a powerful tool to tailor the light–matter interaction for attaining engineered operations against physical constraints in experimental systems, for example, the frequency selectivity required in a molecular qubit system[33] or for coupling two transmon qubits in superconducting qubit system[34].

## RESULTS
### The qubit and consecutive-state transfers

We experimentally implemented the shortcut pulses as described in "Methods" in the $Pr^{3+}$ ion-doped $Y_2SiO_5$ ensemble qubit system. The schematic energy levels of the Pr ions are shown in Fig. 1a, where two ground hyperfine levels define the qubit. $\Omega_s$ and $\Omega_p e^{i\varphi}$ represent Rabi frequencies on the optical transitions from the qubit states to the excited state $|e\rangle$. In this work, we demonstrated high-fidelity operations for two types of transfers. They both start from the initial state $|1\rangle$ but aim for different target states, $|\psi_{tg}\rangle = \sin\theta|0\rangle + \cos\theta\, e^{i\varphi}|1\rangle$. In the first type of transfer, the target state is the qubit level $|0\rangle$ ($\theta = \pi/2$ and $\varphi = 0$). In the other type of transfer, the target state is either one of the four superposition states with $\theta = \pi/4$, and $\varphi = 0, \pi/2, \pi$, or $3\pi/2$. These superposition states are denoted as $|\psi_{tg}^{sup}\rangle$.

For both types of transfer, a sequence of consecutive-state transfers between the initial state, $|1\rangle$, and the target state was carried out in order to determine the transfer fidelity. The qubit initially starts in level $|1\rangle$, as shown in Fig. 1b. Population in level $|1\rangle$ is experimentally characterized by two peaks (peaks 4 and 5 in Fig. 1b) in an absorption spectrum inside a transparent spectral window (TSW), created by optical pumping, which has a frequency width of 18.3 MHz[5]. Each of the peaks has an inhomogeneous frequency width of about 170 kHz (full width at half-maximum, FWHM), and represents the transition from $|1\rangle$ to $|e1\rangle$ or $|1\rangle$ to $|e\rangle$, respectively, as seen in Fig. 1a. The different heights of the peaks are caused by different oscillator strengths of the respective transitions and both peaks can be used to calculate the population in level $|1\rangle$. In the same manner as for the $|1\rangle$ state, the population in the $|0\rangle$ state is also represented by absorption peaks (peaks 1, 2, and 3 in the inset in Fig. 1b), corresponding to transitions from $|0\rangle$ to the three different levels of the excited state.

### Gate operations between $|1\rangle$ and $|0\rangle$

In this section, the gate operation experiment is briefly described and we show how the excitation pulses were characterized. We then explain how the transfer fidelity was obtained from the experimental data and we compare this fidelity with predictions from simulations.

Consecutive population transfers between $|1\rangle$ and $|0\rangle$ were performed up to 18 times in the experiments. A schematic of the pulse sequences is shown in Fig. 2a, where a TSW is created in the first step. Following this, an ensemble of qubit ions is initiated in the $|1\rangle$ state. Then 18 iterations of population transfers are performed. In the Nth ($N = 1, 2, \ldots 18$) iteration, $N$ consecutive population transfers between $|1\rangle$ and $|0\rangle$ are performed, for example, in the third iteration the population is first transferred from $|1\rangle$ to $|0\rangle$, then back to $|1\rangle$, and finally to $|0\rangle$. After the last transfer in each iteration, a weak frequency-scanning readout pulse is applied to measure the absorption spectrum in the TSW. Following each readout, strong frequency-scanning pulses are implemented to optically pump any absorbing ions inside the TSW into a ground-state hyperfine level, the absorption peaks of which lie outside of the TSW. This includes any absorption created through either resonant or nonresonant optical-state transfers or through hyperfine relaxation caused by spontaneous spin flips of ions outside the TSW. These erasing pulses reset the TSW before the next iteration is initiated.

Two different pulses are used for the actual state transfer operation. One aims at transferring $|1\rangle$ to $|0\rangle$, the other aims at transferring the state from $|0\rangle$ back to $|1\rangle$. The first type of pulses is described by Eqs. (11)–(14) in "Methods", where $\theta = \pi/2$, $\varphi = 0$, and the optimized $a_n$ coefficients satisfying Eqs. (15) and (16) are shown as Case 1 in Table 1. The $|0\rangle \to |1\rangle$ pulses are identical to the $|1\rangle \to |0\rangle$ pulses, except that the envelopes and phases of $\Omega_p$ and $\Omega_s$ are interchanged. Intensity envelopes of the $|0\rangle \to |1\rangle$ pulses used in experiments are shown in Fig. 2b (blue curves for $|\Omega_p|^2$, red curves for $|\Omega_s|^2$), which match with the theoretical forms based on Eqs. (11)–(14) (dash-dotted black curves) within ±3% as shown by the inset. The beating generated by the two light fields detected with the reference detector PD1 (as seen in the schematic experimental setup in Fig. 1c) is shown by the red curves in Fig. 2c, which agrees well with the pattern theoretically obtained (dash-dotted black curves). The small discrepancy could be caused by the imperfect calibration on the nonlinear diffraction efficiency of the acousto-optical modulator (AOM) in response to different driving frequencies.

The experimental results of the population transfer between $|1\rangle$ and $|0\rangle$ as a function of the number of transfers $N$ is shown in Fig. 3a, where the blue (red) data points represent the normalized population on $|0\rangle$ ($|1\rangle$) after $N$ transfers. Depending on the parity of $N$, the final population is concentrated on either $|0\rangle$ ($N$ is odd) or $|1\rangle$ ($N$ is even). As $N$ increases, the total operational fidelity decreases because of two facts: (i) decoherence increases with the time that the ions spend in the excited state and this time is directly proportional to the number of transfers performed; (ii) the pulses do not transfer the qubit profile perfectly, partly because of the frequency detuning between the center frequency of the pulse and the individual atomic transition frequencies of the ions caused by the inhomogeneous broadening. This second error can also be expected to accumulate for $N$ transfers. Every data point shown in this work is an average of 100 experimental datasets, and the error bar is the standard deviation. More details on the population and error bar calculations are available in "Methods" (Data evaluation and Credibility of the data evaluation).

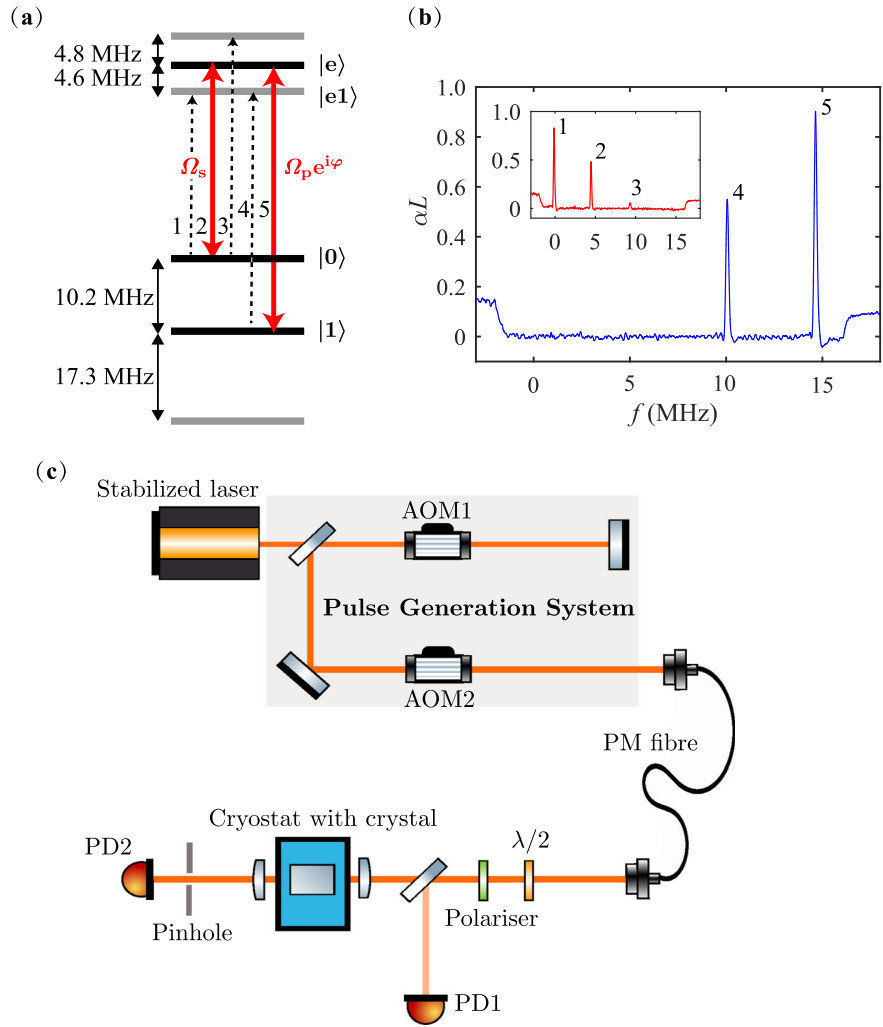

**Fig. 1 Relevant energy levels and the experimental setup. a** The qubit is represented by ground-state levels $|0\rangle$ and $|1\rangle$. They are coupled through optical transitions $|0\rangle \leftrightarrow |e\rangle$ (Rabi frequency $\Omega_s$) and $|1\rangle \leftrightarrow |e\rangle$ (Rabi frequency $\Omega_p$). $\varphi$ is an arbitrary time-independent phase factor applied to $\Omega_p$. The five transitions relevant for measuring the population in the qubit levels are labeled as 1–5. **b** A qubit is initiated at level $|1\rangle$ in a transparent spectral window that has a frequency width of 18.3 MHz. This window covers the frequencies of transitions 1–5 shown in **a**, where the frequency of transition 1 is defined as zero frequency and $\alpha L$ refers to the optical depth. Peaks 4 and 5, located at 10.2 MHz and 14.8 MHz, represents transition 4 and transition 5, respectively. Peak 1–3 located at 0, 4.6, and 9.4 MHz represents the transition 1, 2, and 3, respectively, as shown by the inset. Each transition has an inhomogeneous FWHM linewidth of 170 kHz. **c** The schematic experimental setup. AOM1 (AOM2): the 1st (2nd) acousto-optical modulator. PM fibre: a polarization-maintaining fiber, PD1 (PD2): photodetector1 (photodetector2). More information can be seen in "Methods" (Experimental system).

To evaluate the transfer fidelity $F_{10}$ (from $|1\rangle$ to $|0\rangle$) and $F_{01}$ (from $|0\rangle$ to $|1\rangle$), we investigated how the final fidelity resulting from $N$ transfers in each iteration, $F(N)$, changes with $N$. The result is shown in Supplementary Fig. 1. Here, $F(N)$ is defined as $|\langle\psi_{exp}|\psi_{tg}\rangle|^2$, where $|\psi_{exp}\rangle$ is the qubit state experimentally achieved, and $|\psi_{tg}\rangle$ is the target state. The target state is $|0\rangle$ for all the odd data points (blue points in Fig. 3a) and $|1\rangle$ for even data points (red points in Fig. 3a). Assuming the accumulated transfer fidelity between $|1\rangle$ and $|0\rangle$ is sufficiently high, $F(N)$ depends on $F_{10}$ and $F_{01}$ as

$$F(N) = F_{10}^{(N+1)/2} F_{01}^{(N-1)/2} \text{ for } N = 1, 3, 5, \ldots 17 \quad (1)$$

$$F(N) = F_{10}^{N/2} F_{01}^{N/2} \text{ for } N = 2, 4, 6, \ldots 18 \quad (2)$$

According to our simulation, Eqs. (1) and (2) are good approximations for $N \leq 6$. The drop in fidelity between two neighboring odd or even data points was caused by the two additional transfers between $|1\rangle$ and $|0\rangle$. That means

$$F_{10} \cdot F_{01} = F(N+2)/F(N) \text{ for } N = 1, 2, 3, \ldots 16 \quad (3)$$

Doing this division for all the 18 data points, we eventually get 16 products of $F_{10}$ and $F_{01}$. From these products, we further calculated $F_{10}$ ($F_{01}$) by assuming they are equal ($F_{10} = F_{01} = F$). This is in fact reasonable as the pulses aiming for the transfer from $|1\rangle$ to $|0\rangle$ and vice versa are essentially the same. The results are shown as the blue circles in Fig. 3b. Among these 16 points, we focus on the first four as for higher $N$ ($\geq 6$) Eqs. (1) and (2) are not good approximations any more. The average transfer fidelity $F$ of these four points is $97 \pm 2\%$. The shaded area in Fig. 3b denotes the simulation results with a coherence time between 44 and 132 μs. In ref. [30], the coherence time in a 0.1% $Pr^{3+}:Y_2SiO_5$ crystal was measured as a function of excitation fluence using photon echoes. A π-pulse excitation of a 200 kHz wide ensemble qubit gives an excited-state density of $\sim 2 \times 10^{14}$ excited ions per $cm^3$. Using the data from reference [30], 66 and 132 μs would correspond to excited-state densities of $2 \times 10^{14}$ and zero excited ions per cm³, respectively. In quantum gate operations, the excited-state density may not only affect the fidelity through the coherence time. For example, coherent excitation of two ions, 1 and 2, with the same resonance frequency for the $|1\rangle \leftrightarrow |e\rangle$ transition, but a strong mutual excited-state dipole-dipole interaction, preventing both from being simultaneously excited to the upper state, will create entangled states $(|1\rangle_1|e\rangle_2 + |e\rangle_1|1\rangle_2)$. This is not explicitly modeled in the simulation, but we take height for additional effects like these by also including simulations with a somewhat shorter coherence time, 44 μs. This value is obviously a somewhat arbitrary choice. In ref. [30], this coherence time would correspond to an excited-state density of $4 \times 10^{14}$ ions per cm³. The simulated results agree with experiments within the error bars. More details on the simulation are available in "Methods" (Simulation methods).

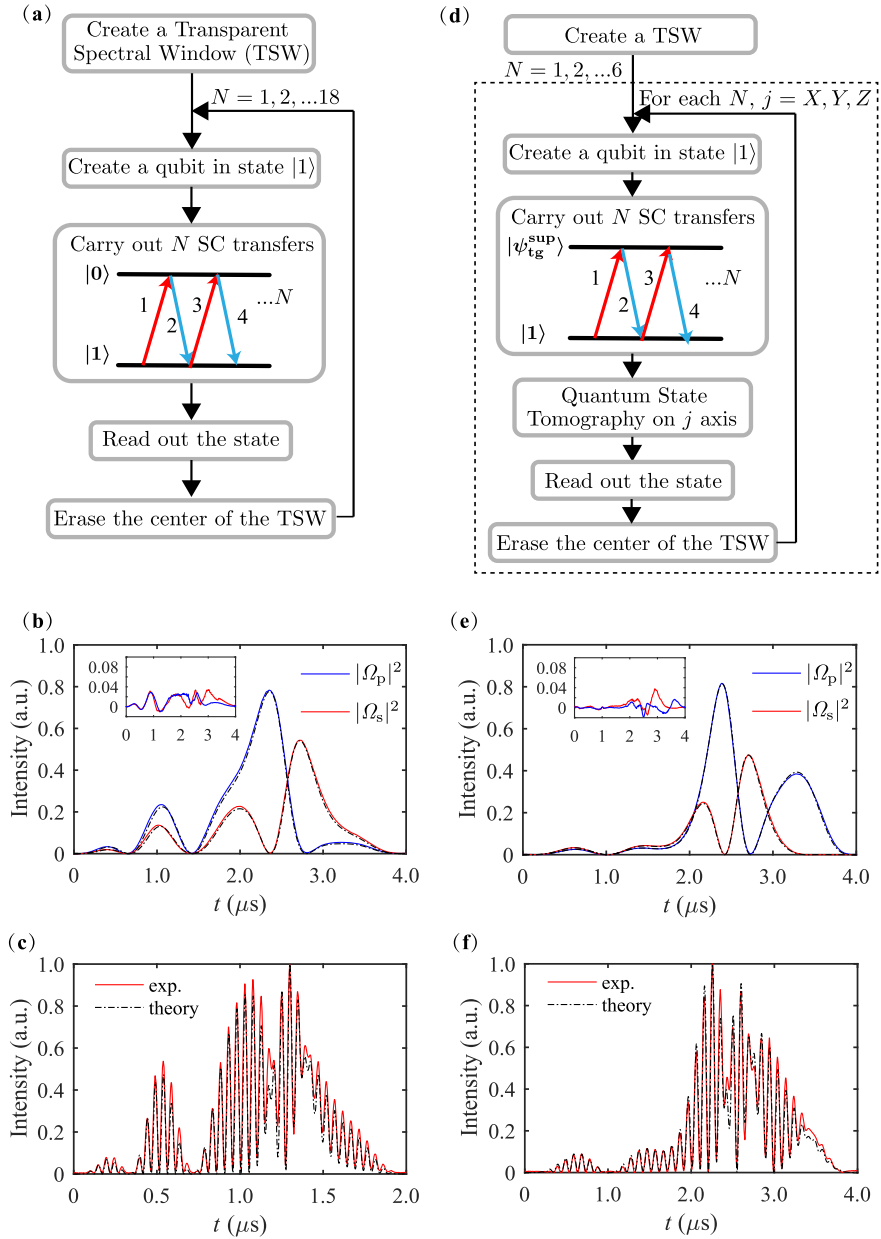

**Fig. 2 Pulses sequences and pulse envelopes.** The left panel shows the figures for the transfer between $|1\rangle$ and $|0\rangle$, and the right panel for the transfer between $|1\rangle$ and $\left|\psi_{tg}^{sup}\right\rangle$ state. **a**, **d** Schematics of the pulse sequence used in experiments, where SC denotes shortcut. **b**, **e** Intensity envelopes of the individual pulses in experiments (solid curves) and in theory (dash-dotted black curves). The blue curve shows $|\Omega_p|^2$ and the red $|\Omega_s|^2$. The difference between the experimentally implemented pulses and the theoretical pulse shapes are shown by the insets. **c**, **f** Beating of the two fields in experiments (red curve) and in theory (dash-dotted black curve).

### Gate operations between $|1\rangle$ and a superposition state

This section largely follows the structure in the $|1\rangle$ to $|0\rangle$ population transfer described in the section above, but it also in detail discusses the quantum state tomography (QST) operations used to characterize the $\left|\psi_{tg}^{sup}\right\rangle$ state.

A sequence of gate operations between $|1\rangle$ and each of the four superposition states $\left(\left|\psi_{tg}^{sup}\right\rangle\right)$ was implemented. The pulse sequence is schematically shown in Fig. 2d, where a difference from the previous case is that QST was implemented right before the readout pulse to characterize the state. To fully

| Table 1. | Optimized $a_n$ values in different cases. | | | | | | | |
|---|---|---|---|---|---|---|---|---|
| Case | $a_1$ | $a_2$ | $a_3$ | $a_4$ | $a_5$ | $a_6$ | $a_7$ | $a_8$ |
| 1 | −0.9911 | −0.5120 | 0.4216 | 0.1530 | 0.0056 | −0.0350 | −0.0431 | −0.0472 |
| 2 | −1.0368 | −0.4374 | 0.2435 | −0.0359 | −0.0008 | 0.0284 | 0.0443 | −0.0190 |
| 3 | −0.9672 | −0.3908 | 0.1210 | 0.1057 | −0.0242 | −0.0625 | 0.1036 | −0.0333 |

Case 1: population transfers between $|1\rangle$ and $|0\rangle$. Case 2: Transfer from $|1\rangle$ to $|\psi_{tg}^{sup}\rangle$, which is any one of the four superposition states, $(|0\rangle \pm |1\rangle)/\sqrt{2}$, $(|0\rangle \pm i|1\rangle)/\sqrt{2}$. The $a_n$ values are the same regardless of the superposition state as the only difference between the pulses for different superposition states is the different constant phase applied to $\Omega_p^{1,sup}$. Case 3: Transfer from $|\psi_{tg}^{sup}\rangle$ to qubit level $|1\rangle$. In each case, $t_f = 4\,\mu s$, and $a_1$ and $a_2$ are the mostly dominant parameters, except for Case 1, where $a_3$ contributes almost equally as much as $a_2$. The $a_1$ ($a_2$) parameters can be altered by about ±5% (±7–15%) without changing the transfer fidelity in the simulation by >1%.

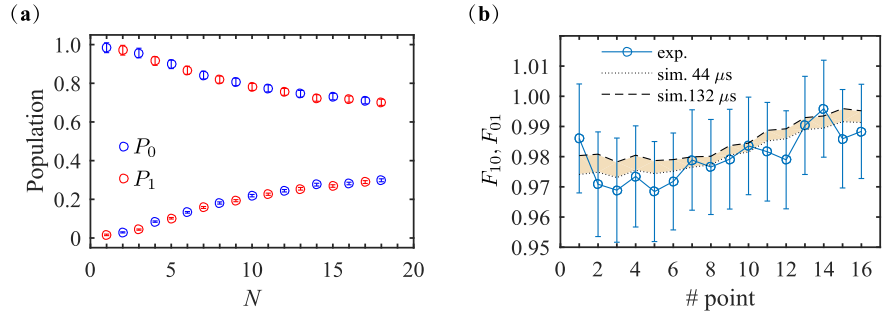

Fig. 3 **Results of population transfers between the two-qubit levels. a** Population distribution as a function of the number of transfers (N) performed in experiments. The error bar is calculated using the 100 experimental data points as described in "Methods" under the Data evaluation part. The same rule applies to all the other figures. **b** The calculated transfer fidelity between $|1\rangle$ and $|0\rangle$ in experiments (blue circles) and simulation (shaded area). The first four data points were used for evaluating the transfer fidelity as Eqs. (1) and (2) apply only for small N (≤6). Simulations were done at two different coherence times, 44 μs and 132 μs.

characterize the final state for each iteration (a specific N), the whole package (sequences inside the dotted rectangle in Fig. 2d) has to be run three times, one time for each tomography axis (X, Y, or Z). Two sets of pulses are required to perform the state transfers. One is for the transfer from $|1\rangle$ to $|\psi_{tg}^{sup}\rangle$, the other is for the backward transfer. Rabi frequencies of the first type of pulses, $\Omega_{p,s}^{1,sup}$, are calculated from Eqs. (11)–(14), and the optimum $a_n$ parameters are shown as Case 2 in Table 1. The second type of pulses are available in two options as stated in ref. [28]. Here, we used the option that the Rabi frequencies, $\Omega_{p,s}^{sup,1}$, drive the Bloch vector rotating backward along the same route as $\Omega_{p,s}^{1,sup}$. This requires that $\Omega_{p,s}^{sup,1}$ are the time reversal of the first transfer $\Omega_{p,s}^{1,sup}$ and, in addition applying a phase shift of π. Thus $\Omega_{p,s}^{sup,1}(t) = -\Omega_{p,s}^{1,sup}(t_f - t)$, where $t_f$ denotes the length of the pulse. For these Rabi frequencies, we re-optimized the pulses parameters $a_n$ to optimize the performance of the transfer. The optimum values are shown as Case 3 in Table 1. Intensity envelopes of the individual pulses used for the transfer from $|1\rangle$ to $|sup\rangle$ in experiments are shown in Fig. 2e (blue curves for $|\Omega_p|^2$, red curves for $|\Omega_s|^2$), which agree with the theoretical envelopes (dash-dotted black curves) within ±4% as shown by the insets. The beating of the two light fields detected with the reference detector PD1 is shown in Fig. 2f (red curves), and agrees well with the theoretical results (dash-dotted black curves).

Although shortcut pulses have the advantage of being much faster and therefore have higher operation fidelity, they only work when the initial state is known and can therefore not be used to characterize an unknown state. Therefore, complex hyperbolic secant pulses were used as tomography pulses. They have previously been characterized with an average QST fidelity of 93%[5]. The tomography fidelity is considerably lower than the shortcut pulse transfer fidelities and need to be compensated for.

The tomography results on the state resulting from ten consecutive transfers between $|1\rangle$ and $(|0\rangle - i|1\rangle)/\sqrt{2}$ are shown in Fig. 4a–c for the X, Y, and Z axis, respectively. The magnitude of each component of the Bloch vector is evaluated by the population difference between state $|0\rangle$ and $|1\rangle$[5]. Both the X and Y components are nearly zero, but the Z component is significant. The reconstructed Bloch vector is close to the −Z axis (i.e. $|1\rangle$) as expected.

However, our simulations show that these QST pulses have errors, and more crucially, the error depends on the exact superposition state, which the QST should characterize. This is a systematic effect occurring because the light pulses targeting the $|0\rangle \leftrightarrow |e\rangle$ and $|1\rangle \leftrightarrow |e\rangle$ transitions also drive other hyperfine transitions from the $|0\rangle$ or $|1\rangle$ to either level in the excited state via off-resonant excitations. This systematic effect yields a symmetric tomography error as a function of initial state as shown in Fig. 4d. For a state on the +X axis the QST fidelity is 95–96%. If we read out a state, 10° clockwise from the +X axis the QST readout fidelity will be ~94%; 45° clockwise relative the +X axis the readout state fidelity is ~90%. For a state rotated clockwise zero degrees (10°, 45°) from the +Y axis we get QST a readout fidelity of 95–96% (~97%, 102%). Note that the readout fidelity of the QST in simulation can be >100% since it is constructed from three separate measurements, and the errors of the readout can align in such a way to yield a Bloch vector that has a magnitude that is larger than one. In brief, the simulations show that the QST fidelity for reading out a single state varies strongly depending on how many degrees, $\vartheta$, the state is rotated relative to the X axis. However, the average of the QST fidelity for the four states all at 90° angle, but each rotated an angle, $\vartheta$, with respect to the +X, +Y, −X, −Y axes, is essentially independent of $\vartheta$. Further, we expect the qubit operation fidelity should essentially be the same for cases rotated 90° because, as mentioned in the text to Table 1, the $a_n$ values are the same regardless of the superposition state, since the only difference between the shortcut pulses for different superposition states is the different constant phase applied to $\Omega_p^{1,sup}$. Therefore, we should be able to average the fidelity results of the four different superposition states, each rotated by 90° relative to one another, to drastically reduce the variation in the QST readout fidelity. This averaging is described in the next section.

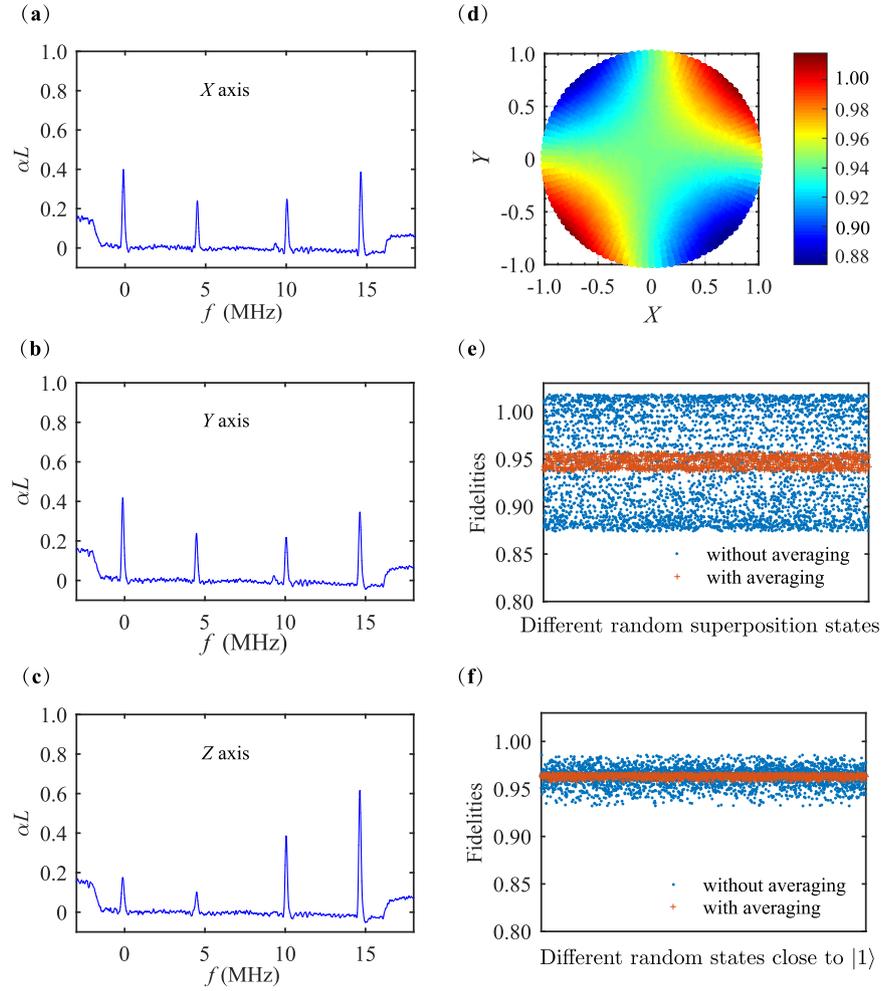

**Fig. 4 Tomography results and theoretical investigation on tomography. a–c** the experimental tomography results on $X$, $Y$, and $Z$ axis of reading out the state after ten transfers between $|1\rangle$ and $(|0\rangle - i|1\rangle)/\sqrt{2}$. **d** The QST fidelity of reading out a superposition state randomly located on a Bloch sphere (top view). **e** QST fidelity of randomly generated superposition states without and with averaging. **f** Same as **e**, but for randomly generated states close to $|1\rangle$.

For the four superposition states which were experimentally investigated ($+X$, $+Y$, $-X$, $-Y$), the fidelity of creating them using the shortcut pulses should actually be the same, as explained in the previous paragraph. Thus, the variation in fidelity observed for these four different superposition states after being read out by the QST might be caused by a state-dependent error in the QST operation. To examine this hypothesis, we carried out a simulation, where we randomly generated 1000 superposition states that lay close to the equator (their $Z$ components were between $-0.2$ and $0.2$), and rotated each of the states by 90, 180, and 270°, respectively, to get three additional superposition states for each original randomly generated state. These four superpositions states mimicked the four experimentally generated superposition states. We then simulated the QST readout on all $1000 \times 4 = 4000$ states, and found that the QST readout fidelity spreads as shown by the blue dots in Fig. 4e. However, if we first performed an average on the fidelities of the four superposition states, then it only varies between 91 and 92%. We did the same analysis on state $|1\rangle$, where we saw the same behavior, as shown in Fig. 4f. However, the variations for reading out state $|1\rangle$ were much less than for the $|sup\rangle$ state, which also agrees with the experimental data, as shown by the inset in Supplementary Fig. 2d.

Based on the investigation of the QST error, we averaged our four experimental results. The overall fidelity $F(N)$ resulting from $N$ transfers between $|1\rangle$ and $|\psi_{tg}^{sup}\rangle$, including both the state transfer and the QST readout fidelity, are shown as the blue circles in Fig. 5a, where the vertical axis is in the logarithm scale and shaded area represents the simulated results assuming the same coherence times as in the transfers between $|1\rangle$ and $|0\rangle$. The individual experimental data for each superposition state is available in Supplementary Fig. 2. Similar to the population transfers in the previous section, $F(N)$ depends on the fidelity of each transfer, $F_{1s}$ (from $|1\rangle$ to $|\psi_{tg}^{sup}\rangle$) and $F_{s1}$ (from $|\psi_{tg}^{sup}\rangle$ to $|1\rangle$), but it now also depends on the QST readout fidelities, $QT_s$ (reading out a superposition state) and $QT_1$ (reading out $|1\rangle$). We obtain

$$F(N) = F_{1s}^{(N+1)/2} \cdot F_{s1}^{(N-1)/2} \cdot QT_s, \quad N = 1, 3, 5 \qquad (4)$$

$$F(N) = F_{1s}^{N/2} \cdot F_{s1}^{N/2} \cdot QT_1, \quad N = 2, 4, 6 \qquad (5)$$

Here, we assume that any error in $QT_s$ and $QT_1$ does not change appreciably between $F(N)$ and $F(N+2)$. This assumption should be even more valid thanks to the reduction in the variation of the QST readout fidelity through averaging the four superposition states as discussed above. The division between $F(N+2)$ and $F(N)$

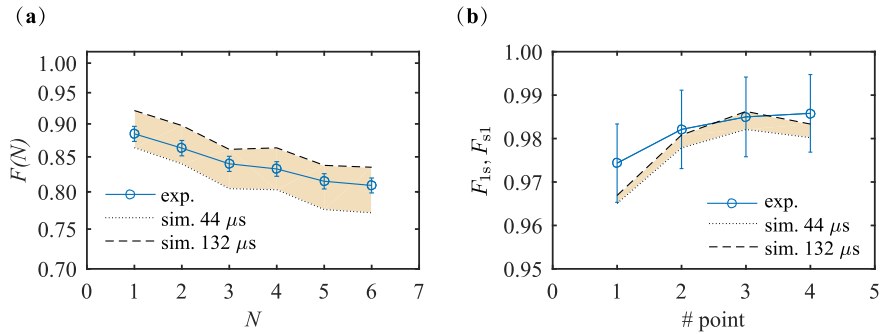

**Fig. 5 Results of transfers between $|1\rangle$ and $\left|\psi_{\text{tg}}^{\text{sup}}\right\rangle$.** **a** The overall fidelity (including the QST readout fidelity) resulting from $N$ transfers between $|1\rangle$ and a $\left|\psi_{\text{tg}}^{\text{sup}}\right\rangle$ as a function of $N$ in experiments (blue circles) and simulation (shaded area) in the logarithm scale. **b** The calculated transfer fidelity between $|1\rangle$ and $|\text{sup}\rangle$ in experiments (blue circles) and simulation (shaded area). Simulations were done at two different coherence times, 44 μs and 132 μs.

then provides the product of $F_{1s}$ and $F_{s1}$ as follows,

$$F_{1s} \cdot F_{s1} = F(N+2)/F(N) \text{ for } N = 1, 2, 3, 4 \quad (6)$$

which is independent on $QT_s$ and $QT_1$. $F_{s1}$ could be potentially different from $F_{1s}$ as its optimal pulse parameters, as shown in Case 3 Table 1, are somewhat different. However, we believe their difference is minor compared with the error bar of the fidelity. Thus, we assume they are the same and the transfer fidelities could be calculated from the product shown in Eq. (6). The results can be seen in Fig. 5b, where the blue circles and orange region denote experimental and simulation results, respectively. The experimental results agree with the simulation results within the error bars. The average of all four experimental data points provides $F_{1s} = F_{s1} = 98 \pm 1\%$.

In addition to the simulation shown in Fig. 5b, we did another simulation where all state transfers are exactly the same but QST was not used in the end of each iteration, as the final state was numerically known. The transfer fidelities resulting from this simulation are shown in Supplementary Fig. 3, which also agree with the experimental results within the error bar. The fact that the simulated gate fidelities agree with experiments in both cases indicates that the QST can be appropriately simulated.

## DISCUSSION

We have demonstrated high-fidelity gate operations between a known initial qubit state and an arbitrary superposition state using nonadiabatic pulses developed from a protocol combining inverse engineering and the optimization of multiple coefficients in the pulse parameters. We have identified and investigated many factors, such as the QST error, limitations in the readout (deconvolution) process, uncertainties in the oscillator strength, effect of the earth magnetic field, etc., which may give uncertainties to the presented fidelity, as shown in "Methods" (Credibility of the data evaluation). However, all uncertainties are within the error bar of the experimental results, and the simulation also agrees with the experimental data within the error bars. Compared with earlier work, we could reduce the error in fidelity from ~7 to <3%, which is made possible by the reduction of the effect of dephasing because the time spent by the ions in an optically excited state is shorter. In the current material, increasing the Rabi frequency may reduce the pulse length and the excited-state decoherence contribution even further, however, higher Rabi frequencies could potentially result in stronger off-resonant excitations. Materials with longer coherence time ($T_2$) and larger ground/excited level separations, for example, $Eu^{3+}:Y_2SiO_5$, where $T_2$ can be 2.6 ms in a magnetic field of 100 G[29] would give better fidelities also for ensemble qubits as the tolerance to both the time spent in the excited state and off-resonant excitations would be larger.

More interesting, although the pulses were implemented on REI ensemble qubits in this work, they could well be used for quantum state engineering on single rare-earth ions. There is presently significant progress in detecting and interacting with individual ions in REI-doped crystals[35–38]. Single-ion operations in rare-earth crystals are projected to enable gate fidelities >99.9%[39]. The scalability of rare-earth single-ion qubit schemes, similar to that in reference[32], is presently investigated within the SQUARE project by the authors. While there is no need for having pulses that compensate for inhomogeneous broadening, high-fidelity operations on a single-ion qubit in such schemes still require that off-resonant excitation of other qubits closeby in both space and frequency is sufficiently low. Because also for single- and multi-qubit operations using single-ion qubits, other qubits which are close in frequency, but not participating in the gate operations should ideally be kept untouched to avoid fidelity loss due to instantaneous spectral diffusion effects[29].

This work shows that the multiple degrees of freedom available in the inverse engineering techniques utilized in this work provide opportunities to handle such demands. In particular, this work indicates that it should be directly applicable for systems where the initial quantum state is known, e.g. for preparing a specific measurement state for quantum sensing applications.

Our demonstration proves that the combination of the inverse engineering technique based on LR invariants and pulse optimization is a versatile scheme to tailor the light–matter interaction in a controlled way for experimental systems with physical imperfections, which in the REI ensemble qubit case are the frequency inhomogeneity within the qubit and unwanted off-resonant excitation outside the qubit frequency interval. Fast and robust pulses, such as those demonstrated in this work, could be used in any frequency-addressed system to initialize ancilla qubits in error correction protocols for fault-tolerant quantum computing.

## METHODS

### Shortcut pulses with multiple degrees of freedom

The Hamiltonian of a three-level Λ system as shown by Fig. 1a with rotating wave approximation in the basis of $|1\rangle$, $|e\rangle$, and $|0\rangle$ reads[28],

$$H(t) = -\frac{\hbar}{2} \begin{bmatrix} 0 & \Omega_p(t)e^{i\varphi} & 0 \\ \Omega_p(t)e^{-i\varphi} & 0 & \Omega_s(t) \\ 0 & \Omega_s(t) & 0 \end{bmatrix}, \quad (7)$$

where $\Omega_i = \frac{\mu_i \cdot E_i}{\hbar}$ ($i = $ p, s) is the Rabi frequency, representing the coupling between the laser and the optical transitions; $\mu_i$ denotes the transition dipole moments, and $E_i$ is the electric field of the laser pulse; and $\varphi$ is a time-independent phase implemented to field $\Omega_p$. The Rabi frequencies in Eq. (7) are to be determined via the LR invariant theory so that $H(t)$ drives a known initial state $|\psi_{in}\rangle$ (without loss of generality we set $|\psi_{in}\rangle = |1\rangle$) to a target state $|\psi_{tg}\rangle = \sin\theta |0\rangle + \cos\theta \, e^{i\varphi}|1\rangle$, where $\theta$ and $\varphi$ are arbitrary angles. LR theory tells that the solution of the Schrödinger equation, $i\hbar \, \partial_t|\psi(t)\rangle = H(t)|\psi(t)\rangle$, can be constructed from a superposition of the different eigenstates, $|\phi_n(t)\rangle$, of the invariant, $I(t)$, as $|\psi(t)\rangle = \sum_n C_n e^{i\alpha_n}|\phi_n(t)\rangle$, where $C_n$ is a time-independent constant determined by boundary conditions, and $\alpha_n = \frac{1}{\hbar}\int_0^t \langle\phi_n(t')|i\hbar \frac{\partial}{\partial t'} - H(t')|\phi_n(t')\rangle$ is the LR phase[16]. The invariant of the Hamiltonian, $I(t)$, satisfying

$$\frac{dI(t)}{dt} = \frac{\partial I}{\partial t} + \frac{1}{i\hbar}[I(t), H(t)] = 0, \quad (8)$$

is as follows[28]

$$I(t) = \frac{\hbar\Omega_0}{2}\begin{pmatrix} 0 & \cos(\gamma)\sin(\beta)e^{i\varphi} & -i\sin(\gamma)e^{i\varphi} \\ \cos(\gamma)\sin(\beta)e^{-i\varphi} & 0 & \cos(\gamma)\cos(\beta) \\ i\sin(\gamma)e^{-i\varphi} & \cos(\gamma)\cos(\beta) & 0 \end{pmatrix}, \quad (9)$$

where $\Omega_0$ is a constant in unit of frequency, $\gamma(t)$ and $\beta(t)$ are time-dependent auxiliary parameters. This invariant has three eigenstates[28], however, in this work we are particularly interested in the eigenstate

$$|\phi_0(t)\rangle = \begin{pmatrix} \cos(\gamma)\cos(\beta)e^{i\varphi} \\ -i\sin(\gamma) \\ -\cos(\gamma)\sin(\beta) \end{pmatrix}, \quad (10)$$

which also itself is a solution of the Schrödinger equation[28]. Therefore, $H(t)$ can drive the state $|\psi(t)\rangle$ evolving along $|\phi_0(t)\rangle$. By imposing the boundary conditions, $\gamma(0) = 0$, $\gamma(t_f) = \pi$, $\beta(0) = 0$, and $\beta(t_f) = \pi - \theta$, we ensure $|\phi_0(t)\rangle$ coincides with the initial and target state of the qubit, i.e. $|\phi_0(0)\rangle = |\psi_{in}\rangle$ and $|\phi_0(t_f)\rangle = |\psi_{tg}\rangle$. These boundary conditions ensure that the target state can be achieved regardless of what ansatz of the auxiliary parameters is used. However, the exact ansatz about the time-dependence of the auxiliary parameters determines the exact dynamics of $|\phi_0(t)\rangle$. This provides us the opportunity to control the state evolution dynamics by choosing the appropriate ansatz of the auxiliary parameters, which are favorable for the experimental system. To this end, we set the auxiliary parameters as follows[28]

$$\gamma(t) = \frac{\pi}{t_f} \cdot t + \sum_{n=1}^{8} a_n \cdot \sin\left(\frac{n\pi}{t_f} \cdot t\right) \quad (11)$$

$$\beta(t) = \frac{\pi - \theta}{2} \cdot [1 - \cos\gamma(t)], \quad (12)$$

where $a_n$ are coefficients of each sinusoidal component. The values of $a_n$ may affect the pulse performance, but have no influence on the boundary conditions of $|\phi_0(0)\rangle$ and $|\phi_0(t_f)\rangle$.

The Rabi frequencies connect with the auxiliary parameters $\gamma(t)$ and $\beta(t)$ through the following auxiliary differential equation based on Eqs. (7)–(9):

$$\Omega_p = -\dot\gamma(t) \cdot [(\pi - \theta) \cdot \cos\gamma(t)\sin\beta(t) + 2\cos\beta(t)] \quad (13)$$

$$\Omega_s = -\dot\gamma(t) \cdot [(\pi - \theta) \cdot \cos\gamma(t)\cos\beta(t) - 2\sin\beta(t)]. \quad (14)$$

For practical reasons, one may prefer $\Omega_i(0) = \Omega_i(t_f) = 0$ to avoid any redundant frequency components in the pulses, which implies that

$$a_1 + 3a_3 + 5a_5 + 7a_7 = 0 \quad (15)$$

$$a_2 + 2a_4 + 3a_6 + 4a_8 = -0.5. \quad (16)$$

Besides the restrictions shown in Eqs. (15) and (16), there are still six degrees of freedom available in $a_n$ ($n = 1, 2, 3, \ldots 8$). We can use this freedom to tailor the light–matter interaction for achieving a high-fidelity operation within a frequency interval designated to the qubit, while simultaneously having a small effect on frequency channels outside of this interval[28]. The optimization for $a_n$ are described in the Simulation methods section shown below. A pulse length of 4 μs is considered in this work. The main pulse duration is purely restricted by the maximal Rabi frequencies experimentally available. The two pulses shown in Eqs. (13) and (14) are generated by an arbitrary waveform generator, which then drives an AOM to create the respective light pulses with appropriate driving frequencies, durations and phases. A description of the exact experimental system can be seen in the Experimental system section below.

### Material

A $Pr^{3+}$ doped $Y_2SiO_5$ crystal, with a doping concentration of 0.05% was used for all experiments. The sample had dimensions $10 \times 10 \times 0.8$ mm, along the $D_1 \times D_2 \times C_2$ axes, respectively. The experiments were performed on the transition between the lowest crystal-field levels of the $^3H_4$-$^1D_2$ (Site 1) at 606 nm. Light was propagated along the $C_2$ axis, with its polarization axis along $D_2$. The lifetime $T_1$ of the excited state $^1D_2$ is 164 μs and the intrinsic coherence time, $T_2$, has been measured to be 132 μs without a magnetic field applied[30]. The inhomogeneous linewidth was ~10 GHz and the absorption coefficient ($a$) of the crystal along the $D_2$ direction was ~4000 m$^{-1}$.

### Experimental system

The schematic experimental setup is shown in Fig. 1c. The light source is a ring dye laser (Coherent 699-21), where the dye being used was Rhodamine 6 G mixed in ethylene glycol. It was pumped by a Nd:YVO$_4$ laser (Coherent Verdi), resulting in an output power of ~500 mW. The linewidth of the laser light is reduced to a few tens of Hz using the Pound–Drever–Hall technique by locking to an ultralow thermal expansion cavity. The pulse generation system consisted of two AOMs: (i) AOM1 (double pass, AA.ST.200/B100/A0.5-vis) is used for generating all the pulses used for qubit initialization and readout; (ii) AOM2 (single pass, 60 MHz, Isomet 1205C-2) is used in the generation and shaping of the two frequency components of the shortcut pulses (targeting the two optical transitions $|0\rangle$-$|e\rangle$ and $|1\rangle$-$|e\rangle$). Both AOMs were controlled using an arbitrary waveform generator (Tektronix AWG520). A polarization-maintaining (PM) fibre was used to couple light from the dye laser table to another optical table housing the cryostat, while also serving as a spatial filter. The sample was immersed in liquid helium, which was pumped to 2.17 K. No external magnetic field was applied for any of the measurements.

The laser power on the cryostat table was about 50 mW, a part of which was sampled onto reference detector PD1 (Thorlabs PDB150A). A half-wave plate and a polarizer were used to align the polarization axis along the $D_2$ axis of the crystal. This beam was focused onto the crystal down to a spot measuring 70 μm in $1/e^2$ diameter. Using a lens system (275 mm and 750 mm) and a 25-μm-diameter pinhole, the center of the focal spot (across which the intensity variation was <10%) was imaged. The detection system consisted of a photodiode (PD2, Hamamatsu S5973-02), a current amplifier (Femto DHPCA-100) and an oscilloscope (Wave Runner HRO 66Zi) for data acquisition.

### Simulation methods

Simulations performed in this work are based on evolving the Lindblad master equation using the ode45 function in MATLAB, with local and absolute tolerances of $10^{-6}$ which gives a global error of the simulation of around ten times that. The system being simulated contains six quantum levels representing the three ground and three excited hyperfine levels of Pr as seen in Fig. 1a. The incoming laser pulses drive all optical transitions simultaneously with a strength determined by the relative oscillator strengths[40], and detuning based on the energy separation between the hyperfine levels. It also includes the effect of the frequency inhomogeneity of the Pr ions, thus incorporating the effect of the qubit peak shape, which is obtained from the experimental readout. However, it does not include the inhomogeneity of the hyperfine splitting (~30 kHz). The optical lifetime is set to 164 μs. The effect of instantaneous spectral diffusion has been taken into account by adjusting the optical coherence time to fit the experimental data. The reason for using a fitting procedure to adjust the coherence time is that it is unclear how the strong consecutive pulses used for gate operations during the experiments will affect the coherence times. The ground-state hyperfine coherence time is set to 500 μs[41]. The hyperfine relaxation between the ground states, however, has not been considered since that is occurring on a much longer timescale.

### Optimization of pulse parameters

To find the optimum pulse parameters, the $a_n$ in Eq. (11) and constrained by Eqs. (15) and (16), we carried out simulations in the six-level model. Thus, all levels which might be populated by off-resonant excitation in the experiments were included. These parameters were optimized using the simulator described in the Simulation methods section. The optimization algorithms used were existing MATLAB functions based on simulated annealing (simulannealbnd) followed by the fminsearch algorithm. The score improved with increased fidelity for ions within a ±500 kHz frequency region around the center of the qubit peak, but became worse with any change in population outside the TSW in order to minimize off-resonant excitation caused by the pulse. More precisely, the score was calculated by taking the square of the difference between the density matrix obtained from the simulations and the desired density matrix.

### Data evaluation
Every experimental data point presented in each figure is based on 100 consecutively recorded transmission spectra inside a TSW. They were deconvolved into absorption spectra[42] (Fig. 1b), from which the population distribution resulting from the state transfers and further the transfer fidelity was evaluated based on the area of all absorption peaks. For the purpose of decreasing the noise level while avoiding to smear out the signal due to systematic changes in the signal strength due to drift in laser power for the long time period, these 100 absorption spectra were divided into five groups. We evaluated the area of each peak inside the TSW as follows. (i) We did an average over the 20 spectra in each group; (ii) we made a Gaussian fit on each peak (peak 1, 2, 3, 4, and 5, see Fig. 1b) in the averaged absorption spectrum and calculated the respective area under each fit as the peak area ($PA_{i\_}fit_j$, $i, j = 1, 2, … 5$); (iii) we made an average over the results of the five fits, i.e. $PA_i = \text{mean}(PA_{i\_}fit_j)$, as a result we got the five peak areas in each data point. The error bar of every peak area ($PA_{i\_}fit_j\_std$) in each group is calculated from the 68% confidence interval of the fitting parameters in the Gaussian function, and the final error bar of that peak area ($PA_i\_std$) is the average of the five error bars, i.e. $PA_i\_std = \Sigma_j PA_{i\_}fit_{j\_}std/5$. These error bars were propagated to the population and fidelity calculation presented in this work following the error propagation rules.

Based on all the peak areas ($PA_i$), the population distribution in level $|0\rangle$ and $|1\rangle$ are evaluated as follows. The normalized population in level $|0\rangle$ was evaluated from the peak 1 and peak 2 as $P_0 = (PA_1/f_{0,e1} + PA_2/f_{0,e})/(2P) \equiv P'/P$, and in level $|1\rangle$ was evaluated from the peak 4 and peak 5 as $P_1 = (PA_4/f_{1,e1} + PA_5/f_{1,e})/(2P) \equiv P''/P$, where $f_{l,m}$ denotes the oscillator strength of the transition between level $|l\rangle$ and level $|m\rangle$, $P$ represents the total population in both $|1\rangle$ and $|0\rangle$, and $P = P' + P''$. Tomography data were evaluated from the normalized population distribution resulting from each tomography axis in the same way as that previously presented[5].

### Credibility of the data evaluation
We have examined a few factors, which could give minor uncertainties on the fidelities presented in this work. However, the magnitudes of the uncertainties are all well below the given error bar.

(1) The pulse area of the readout pulses used for measuring the transmission spectrum is ~1.7% of a π pulse in this work. This causes a corresponding error in the inferred population of 0.3%, which is well within the error bar about ±2% of our population evaluation, as seen in Fig. 3a.

(2) A deconvolution algorithm[42] is employed to recover the absorption spectrum from the transmission signal. It is designed for low absorption and introduces an error in higher $\alpha L$ values by up to 7–13%. This error has been corrected in the data evaluation. In principle, this correction was not necessary since we found that the effect of the correction on the fidelity is well within the error bars. The reason for this is that the peak areas drop very slowly with the increasing number of transfers. Consequently, two neighboring data points may both shift up or down slightly, but the transfer fidelity is determined by the ratio of these two points and this ratio is almost unaffected by the deconvolution uncertainty.

As a side effect of the error in the deconvolution discussed above, the height of high $\alpha L$ peaks are rescaled. Correspondingly, the oscillator strengths ($f_{l,m}$) used here were slightly rescaled (by up to 8%) from our previously published values[40].

(3) The negative feature on the right edge of the peak centered at 14.8 MHz in Fig. 1b is an artifact of the deconvolution program since it is intended for low $\alpha L$ values. These artifacts modify the qubit peak area slightly. Thus, the change in fidelity resulting from the observed artifact is well within the error bar. To show this we: first, did a Gaussian curve fit to calculate the peak area, which is less affected by the artifacts than a direct integration; secondly, calculated the fidelity from the ratio between two neighboring data points, and the artifacts often largely affect the peak areas equally.

### Systematic errors
We have investigated and corrected the effect of some systematic errors on our experimental results.

(1) In the experiments, there is a wait time of one millisecond right before the readout after the tomography pulses in each iteration. Since the excited-state lifetime is $T_1 = 164$ μs[30], we lose information about the population in the excited state due to decay during this wait time. To estimate this effect, we experimentally measured the excited-state population at 100 μs after 12 transfers had been performed. The result is 6.5% (±2%), which indicates a population of 12% (±4%) in the excited state right after all transfers, i.e., an average of ~1% from each pulse operation. Since the error bar of the data points in Fig. 3a are about ±2%, it would have been difficult to estimate the excited-state population after a few transfers if no wait time had been used, especially since the analysis would become more complicated. However, any population in the excited state after N transfers do affect the fidelity of the following pulses, thus leading to a lower transfer fidelity. Still the excited-state population is limited in the first transfers and we believe additional consideration of the effect of the excited-state population would not significantly alter our results.

(2) The photodetector used to detect the transmitted light, PD2 in Fig. 1c, needs to have sufficient bandwidth to capture the signal changes, as well as sufficient gain to accomplish a good signal-to-noise ratio. The latter requirement is especially critical when a pinhole with a power throughput of ~6% is used in the detection system. However, high gain is accompanied by low bandwidth. Thus, there is a tradeoff between them. In experiments, the response time of PD2 is too slow to fully capture the steep signal changes as the readout pulse is frequency chirped across the absorption peaks. The signal has been corrected for the limited frequency response of the photodetector.

(3) The earth magnetic field (measured to be about 50 μT) in our lab is not compensated for in this work. It could lift the degeneracy in the levels, producing a splitting of about 3 kHz for $|0\rangle$, $|1\rangle$, and $|e\rangle$ states[43]. This splitting leads to a maximal splitting of about 10 kHz in the absorption peaks, which is about 6% of the linewidth of the peak representing the qubit. The effect of the splitting on our qubit state should be negligible.

### DATA AVAILABILITY
All data needed to evaluate the conclusions in the article are presented in the article and/or the Supplementary Figures. Additional data related to this paper may be requested from the corresponding authors.

### CODE AVAILABILITY
The code that supports the findings of this work is available from the corresponding authors upon reasonable request.

Received: 15 January 2021; Accepted: 12 August 2021;
Published online: 14 September 2021

## ACKNOWLEDGEMENTS
We acknowledge the support from National Natural Science Foundation of China (NSFC) (61505133, 61674112, 62074107); Natural Science Foundation of Jiang Su Province (BK20150308); The International Cooperation and Exchange of the National Natural Science Foundation of China NSFC-STINT (61811530020); S.K. acknowledges the support from the Swedish Research Council (no. 2016-04375, no. 2019-04949), the Knut and Alice Wallenberg Foundation (KAW2016.0081) and Wallenberg Center for Quantum Technology (WACQT) (KAW2017.0449); European Union's Horizon 2020 research and innovation program (712721); NanOQ Tech and the Lund Laser Centre (LLC) through a project grant under the Lund Linneaus environment. This project has received funding from the European Union's Horizon 2020 research and innovation program under grant agreement no. 820391 (SQUARE) and no. 654148 Laserlab-Europe. A.W. acknowledges the support from the Swedish Research Council (no. 2015-03989). L.R. acknowledges the support from the Swedish Research Council (no. 2016-05121). X.C. acknowledges the support by the Spanish Ministry of Science and the European Regional Development Fund through PGC2018-101355-B-I00 (MCIU/AEI/FEDER, UE) and the Basque Government through Grant No. IT986-16, the EU FET Open Grant Quromorphic (Grant No. 828826), and EPIQUS (Grant No. 899368) and the Ramon y Cajal program (Grant No. RYC-2017-22482).

## AUTHOR CONTRIBUTIONS
Y.Y., C.S., H.S., and A.K. performed the experiments with contribution from S. Horvath. A.K. performed the simulation and established the experimental MATLAB codes. Y.Y., C.S., A.K., and H.S. evaluated the data. Y.Y., H.S., A.K., and A.W. wrote the article. All authors contributed to the planning of the experiments and commented on the article. The project is supervised by Prof. X.C. and Prof. S.K.

## COMPETING INTERESTS
The authors declare no competing interests.

## ADDITIONAL INFORMATION
**Supplementary information** The online version contains supplementary material available at https://doi.org/10.1038/s41534-021-00473-4.

**Correspondence** and requests for materials should be addressed to Ying Yan, Xi Chen or Stefan Kröll

**Reprints and permission information** is available at http://www.nature.com/reprints

**Publisher's note** Springer Nature remains neutral with regard to jurisdictional claims in published maps and institutional affiliations.